# Properties and Application of Nondeterministic Quantum Query Algorithms

Alina Dubrovska

Department of Computer Science, University of Latvia,
Raina bulv. 29, LV-1459, Riga, Latvia
alina.dubrovska@gmail.com

**Abstract.** Many quantum algorithms can be analyzed in a query model to compute Boolean functions where input is given by a black box. As in the classical version of decision trees, different kinds of quantum query algorithms are possible: exact, zero-error, bounded-error and even nondeterministic. In this paper, we study the latter class of algorithms. We introduce a fresh notion in addition to already studied nondeterministic algorithms and introduce dual nondeterministic quantum query algorithms. We examine properties of such algorithms and prove relations with exact and nondeterministic quantum query algorithm complexity. As a result and as an example of the application of discovered properties, we demonstrate a gap of $n$ vs. 2 between classical deterministic and dual nondeterministic quantum query complexity for a specific Boolean function.

## 1 Introduction

We study the query model to compute Boolean functions, where the input $(x_1, x_2, \ldots, x_n)$ is contained in a black box and can be accessed by asking questions about the values of $x_i$. The goal is to compute the value of the function. The complexity of a query algorithm is determined by the maximum number of questions that it asks. The classical version of this model is known as *decision trees* [2].

Quantum query algorithms can solve certain problems faster than classical algorithms. The best known example of an efficient exact quantum query algorithm is the algorithm for the $PARITY_n$ function: the value is 1 if the input contains an even number of 1. This function can be computed with $\lceil n/2 \rceil$ queries, while any classical algorithm requires $n$ queries.

The theory of computation studies various models: deterministic, nondeterministic, probabilistic and quantum. As in the classical version of decision trees, different kinds of quantum query algorithms are possible: exact, with bounded error or nondeterministic.

The main aim of this research was to study and examine the notion of nondeterministic quantum query algorithms. For quantum nondeterministic query algorithms, de Wolf [7] has proved that it is possible to compute $OR(X) = x_0 \vee x_1 \ldots \vee x_{n-1}$ with 1 question for all $n$, though it is known that the best deterministic algorithm requires all $n$ questions. This is the largest possible gap between complexities of two different kinds of algorithms allowed by a model.

## 2. Notation and Definitions

Let $f(x_1, x_2, ..., x_n) : \{0,1\}^n \rightarrow \{0,1\}$ be a Boolean function. We use $\oplus$ to denote XOR (exclusive OR). We use $\bar{f}$ for the function $1 - f$.

### 2.1. Quantum computing

We apply the basic model of quantum computing. For more details, see textbooks by Gruska [4] and Nielsen and Chuang [5].

An $n$-dimensional quantum pure state is a vector $|\psi\rangle \in C^n$ of norm 1. Let $|0\rangle, |1\rangle, ..., |n-1\rangle$ be an orthonormal basis for $C^n$. Then, any state can be expressed as $|\psi\rangle = \sum_{i=0}^{n-1} a_i |i\rangle$ for some $a_0 \in C$, $a_1 \in C$, ..., $a_{n-1} \in C$. Since the norm of $|\psi\rangle$ is 1, we have $\sum_{i=0}^{n-1} |a_i|^2 = 1$. States $|0\rangle, |1\rangle, ..., |n-1\rangle$ are called *basic states*. Any state of the form $\sum_{i=0}^{n-1} a_i |i\rangle$ is called a *superposition* of $|0\rangle, |1\rangle, ..., |n-1\rangle$. The coefficient $a_i$ is called an *amplitude* of $|i\rangle$.

The state of a system can be changed using *unitary transformations*. Unitary transformation $U$ is a linear transformation on $C^n$ that maps vectors of unit norm to vectors of unit norm. If, before applying $U$, the system was in state $|\psi\rangle$, then the state after the transformation is $U|\psi\rangle$.

The simplest case of quantum measurement is used in our model. It is the full measurement in the computation basis. Performing this measurement on state $|\psi\rangle = a_1|0\rangle + ... a_k|k\rangle$ gives the outcome $i$ with probability $|a_i|^2$. The measurement changes the state of the system to $|i\rangle$ and destroys the original state, $|\psi\rangle$.

### 2.2. Query model

Query model is the simplest model for computing Boolean functions. In this model, the input $x_1, x_2, ..., x_n$ is contained in a black box and can be accessed by asking questions about the values of $x_i$. Query algorithm must be able to determine the value of a function correctly for arbitrary input contained in a black box. The complexity of the algorithm is measured by the number of queries to the black box that it uses. The classical version of this model is known as *decision trees*. For details, see the survey by Buhrman and de Wolf [2].

We consider computing Boolean functions in the quantum query model. For more details, see the survey by Ambainis [1] and textbooks by Gruska [4] and de Wolf [6]. A quantum computation with $T$ queries is a sequence of unitary transformations:

$$U_0 \rightarrow Q \rightarrow U_1 \rightarrow Q \rightarrow ... \rightarrow U_{T-1} \rightarrow Q \rightarrow U_T$$

$U_i$'s can be arbitrary unitary transformations that do not depend on the input bits $x_1, x_2, ..., x_n$. $Q$'s are query transformations. The computation starts with a state $|\vec{0}\rangle$. Then we apply $U_0, Q, ..., Q, U_T$ and measure the final state.

There are several different, but equally acceptable ways to define quantum query algorithms. The most important consideration is to choose an appropriate definition for the query black box, defining the form for asking questions and receiving answers from the oracle.

Next we will precisely describe the full process of quantum query algorithm definition and notation used in this paper.

Each quantum query algorithm is characterized by the following parameters:

1) *Unitary transformations*

All unitary transformations and the sequence of their application (including the query transformation parts) should be specified.

Each unitary transformation is a unitary matrix. Here is an example of an algorithm sequence specification with *T* queries:

$$|\vec{0}\rangle \to U_0 \to Q_1 \to ... \to Q_T \to U_N \to [M],$$

2) *Queries*

To specify a question, we must assign a number of queried variable to each amplitude. Assume we have a quantum state with *m* amplitudes $|\psi\rangle = (\alpha_1, \alpha_2, ..., \alpha_m)$. For the *n* argument function, we define a query as $Q_i = (\alpha_1 \equiv k_1, ..., \alpha_m \equiv k_m)$, where *i* is the number of question, and $k_j \in \{1..n\}$ is the number of queried variable. If $x_{k_j} = 1$, a query will change the sign of the *j*-th amplitude to the opposite sign; in all other cases, the sign will remain as-is.

3) *Measurement*

Each amplitude of a final quantum state corresponds to the algorithm output. We assign a value of a function to each output. We denote it as $M = (\alpha_1 \equiv k_1, ..., \alpha_m \equiv k_m)$, where $k_i \in \{0,1\}$. The result of running an algorithm on input *X* is *j* with a probability that equals the sum of squares of all amplitudes, which corresponds to outputs with value *j*. We denote the probability of obtaining result 0 with $p(f(X) = 0)$, and the probability of obtaining result 1 with $p(f(X) = 1)$.

The following diagram represents the query algorithm in general form:

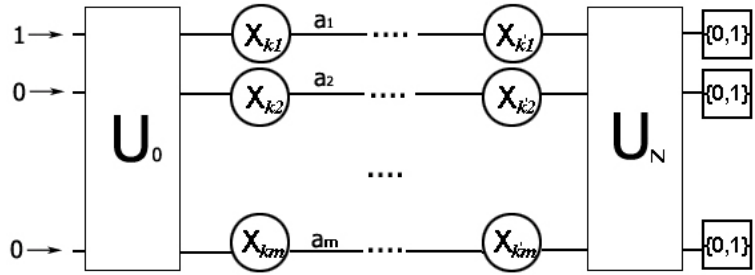

Fig. 1 Graphical representation of a quantum query algorithm.

### 2.3. Query complexity

The complexity of a query algorithm is based on the number of questions it uses to determine the value of a function on worst-case input.

The *deterministic decision tree complexity* of a function *f*, denoted by *D(f)*, is the minimum number of queries that must be performed on any input by an optimal deterministic algorithm for *f* [2].

For deterministic query complexity estimation for a function, the notion of sensitivity *s(f)* is useful. The sensitivity of *f* on input $(x_1, x_2, \ldots, x_n)$ is the number of variables $x_i$ with the following property: $f(x_1, \ldots, x_i, \ldots, x_n) \neq f(x_1, \ldots, 1-x_i, \ldots, x_n)$. The sensitivity of *f* is the maximum sensitivity of all possible inputs. It has been proved in [2] that $D(f) \geq s(f)$.

A quantum query algorithm *computes f exactly* if the output equals *f(x)* with a probability of 1, for all $x \in \{0,1\}^n$. $Q_E(f)$ denotes the number of queries of an optimal exact quantum query algorithm for a function *f* [2].

### 2.4. Nondeterministic query algorithms

Nondeterministic quantum query algorithms were examined by de Wolf in [7]. A *nondeterministic quantum query algorithm* for *f* is defined to be a quantum algorithm that outputs 1 with positive probability if *f(x)*=1 and that always outputs 0 if *f(x)*=0. $NQ_1(f)$ denotes the query complexity of an optimal nondeterministic quantum algorithm for *f*.

We introduce the notion of *dual nondeterministic quantum query algorithm* and study the relations between exact, nondeterministic and dual nondeterministic quantum query algorithm complexity.

**Definition 1** *A dual nondeterministic quantum query algorithm for f is defined to be a quantum algorithm that outputs 0 with positive probability if f(x)=0 and that always outputs 1 if f(x)=1.*

$NQ_0(f)$ denotes the query complexity of an optimal dual nondeterministic quantum algorithm for *f*.

## 3. Properties

In this section, we discuss and prove some properties of nondeterministic quantum query algorithms. First, we describe the relation between complexities of nondeterministic and dual nondeterministic quantum query algorithm.

**Lemma 1** *Nondeterministic quantum query algorithm for a function f can be transformed in a dual nondeterministic algorithm for a function $\bar{f}$ by replacing assigned values for each output $j \in \{0,1\}$ with $(1-j)$. The same is true for transforming algorithms in the opposite direction.*

**Proof.** Let A1 be a dual nondeterministic algorithm for a function *f(X)* with complexity $NQ_0(A1) = k$.

We run algorithm on all inputs and, depending on the result, divide them into three sets:

$A = \{X \mid p(f(X) = 1) = 1\}$

$B = \{X \mid p(f(X) = 0) = 1\}$

$C = \{X \mid p(f(X) = 0) > 0 \ \& \ p(f(X) = 1) > 0 \ \& \ p(f(X) = 0) + p(f(X) = 1) = 1\}$

According to the definition of a dual nondeterministic algorithm, we can assign to each set the result value of running $A1$ on pertinent input (Table 1).

Table 1. Results of running algorithm A1

| $X$ belongs to set: | result of $A1$ |
|---|---|
| A | 1 |
| B | 0 |
| C | 0 |

Table 2. Results of running algorithm A1'

| $X$ belongs to set: | result of $A1'$ |
|---|---|
| A | 0 |
| B | 1 |
| C | 1 |

Now we change the value assignment for each output to the opposite: $0 \rightarrow 1$ and $1 \rightarrow 0$. We denote a new algorithm with $A1'$ and prove that this is a correct nondeterministic algorithm for $\bar{f}$.

We examine the same input sets *A*, *B* and *C* after running A1' and obtain opposite probabilities for sets A and B:

$A = \{X \mid p(f(X) = 0) = 1\}$

$B = \{X \mid p(f(X) = 1) = 1\}$

$C = \{X \mid p(f(X) = 0) > 0 \ \& \ p(f(X) = 1) > 0 \ \& \ p(f(X) = 0) + p(f(X) = 1) = 1\}$

After evaluating the function value obtained by running $A1'$ according to the nondeterministic algorithm definition, we obtain the results represented in Table 2.

When we compare derived results with values from Table 1 we conclude that A1' computes $\bar{f}$ as a nondeterministic algorithm.

The proof in the opposite direction is similar. □

**Theorem 1** *For an arbitrary Boolean function f,* $NQ_0(f) = NQ_1(\bar{f})$.

**Proof.** Follow from Lemma 1. We can take the best existing dual nondeterministic algorithm for function *f* and easily transform it into a nondeterministic algorithm for $\bar{f}$. We change only the value assignment to outputs; the number of questions $NQ_0(f)$ remains the same. In the other direction, we can take the best existing nondeterministic algorithm for $\bar{f}$ and transform it into a dual nondeterministic for $\bar{\bar{f}} \equiv f$ staying with the same $NQ_1(\bar{f})$ queries. □

Next we consider some composite functions and demonstrate a way to use exact quantum query algorithm for a function to construct a nondeterministic quantum algorithm for a more complex function.

The first construction is the composite function MULTI_AND. We denote:

$MULTI\_AND_m(x_1,...,x_{mn}) = \underbrace{f_n(x_1,...,x_n) \wedge f_n(x_{n+1},...,x_{2n}) \wedge ... \wedge f_n(x_{(m-1)n+1},...,x_{mn})}_{m}$

We call the $f_n(x_1,...,x_n)$ base function or sub-function. The composite function MULTI_AND is obtained using a base function structure as a pattern, joining several similar variable blocks with a logical AND operation.

The second construction is a composite function MULTI_OR. We denote:

$$MULTI\_OR_m(x_1,...,x_{mn}) = \underbrace{f_n(x_1,...,x_n) \vee f_n(x_{n+1},...,x_{2n}) \vee ... \vee f_n(x_{(m-1)n+1},...,x_{mn})}_{m}$$

**Theorem 2** *Let $Q_1$ be an exact quantum query algorithm that computes a Boolean function f with k queries. Then a dual nondeterministic quantum query algorithm $Q_2$ exists, which computes function $MULTI\_AND_m(f)$ with the same k queries for all m.*

**Proof.** Let $Q1$ be an exact quantum algorithm with $k$ queries for an $n$-variable function $f_n$. We denote the assignment of values for outputs with $M = (q_1 \equiv k_1, q_2 \equiv k_2,..., q_h \equiv k_h)$, where $h$ is a number of amplitudes.

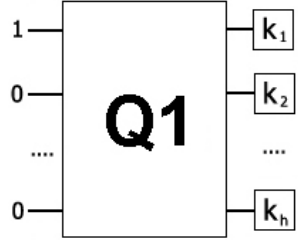

Fig. 2 Exact quantum algorithm Q1 to compute $f_n$.

Now we will construct a quantum algorithm $Q2$, which will compute $MULTI\_AND_m(f)$ as a dual nondeterministic algorithm.

We use the following notation:

$MULTI\_AND_m(X) = f_n(X_1) \wedge f_n(X_2) \wedge ... \wedge f_n(X_m)$,

where $X = X_1 X_2 X_3 ... X_m$,

$\forall i \in \{1..m\}: X_i = (x_{i1} = \alpha_{i1}, x_{i2} = \alpha_{i2},..., x_{in} = \alpha_{in})$ and $\alpha_{ij} \in \{0,1\}$, it is allowed that $X_i = X_j$, even if $i \neq j$.

To evaluate the value for each of $m$ occurrences of $f_n$ we run $Q1$ in parallel. To retain the total sum of squares of amplitudes equal with 1, we separate the initial amplitudes distribution between all $m$ parts of $Q2$. All $Q1$ transformations are unitary and from the structure of the algorithm it follows that $Q2$ transformations also are unitary. Algorithm $Q2$ is demonstrated in Figure 3.

We denote the sum of squares of all amplitudes where output value is "0" in the part of $Q2$ corresponding to $f(X_i)$ with $P_{f(X_i)}("0")$. In a similar manner with $P_{f(X_i)}("1")$, we denote the sum of squares of all amplitudes of $f(X_i)$ parts corresponding to outputs with assigned "1".

If, after running $Q1$ on some input $X_i$ we have a final distribution of amplitudes $(b_{i1}, b_{i2},..., b_{ih})$, then when computing $f_n(X_1) \wedge f_n(X_2) \wedge ... \wedge f_n(X_m)$ with $Q2$, we will obtain the following final distribution of amplitudes:

$$\left((b'_{11}, b'_{12},..., b'_{1h}),(b'_{21}, b'_{22},..., b'_{2h}),...,(b'_{m1}, b'_{m2},..., b'_{mh}),0,...,0\right), \text{ where } b'_{ij} = \frac{1}{\sqrt{m}} b_{ij}.$$

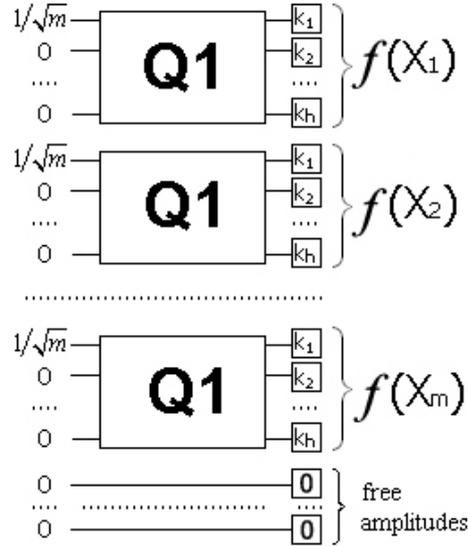

Fig. 3 Quantum query algorithm Q2.

**Lemma 2** *For an arbitrary input $X_i$:*

- *If the result of running Q1 is $f(X_i)=0$, then for Q2 we have $P_{f(X_i)}("0") = \frac{1}{m}$ and $P_{f(X_i)}("1") = 0$.*
- *If the result of running Q1 is $f(X_i)=1$, then for Q2 we have $P_{f(X_i)}("0") = 0$ and $P_{f(X_i)}("1") = \frac{1}{m}$.*

**Proof.** Follows from the properties of exact quantum query algorithm and the fact that the value assignment for outputs in each part of Q2 is the same as in Q1. □

For the entire algorithm Q2 we have:

- If there exists at least one such $i \in \{1..m\}$ that $f(X_i) = 0$, then by Lemma 2 in a part of Q2 corresponding to $f(X_i)$ there will be $P_{f(X_i)}("0") = \frac{1}{m}$. The total probability of obtaining result "0" will be $p(f(X) = 0) > 0$ and in the view of dual nondeterministic algorithm definition we have $MULTI\_AND(X) = 0$.

- If for all $i \in \{1..m\}$ it is true that $f(X_i) = 1$, then in all Q2 parts $P_{f(X_i)}("1") = \frac{1}{m}$ and the total probability is $p(f(X) = 1) = 1$, so in the view of a dual nondeterministic algorithm definition we have $MULTI\_AND(X) = 1$.

That completely agrees with the essence of function MULTI_AND; hence, Q2 computes this function as a dual nondeterministic algorithm. □

The next theorem can be used for complexity estimation.

**Theorem 3** *For an arbitrary Boolean function f, $NQ_0(MULTI\_AND_m(f)) \leq Q_E(f)$.*

**Proof.** Using the approach from Theorem 2, we always can construct a dual nondeterministic algorithm for $MULTI\_AND_m(f)$ based on the best existing exact algorithm for *f*. In this case, we get equality of complexities. It may be possible to find a better algorithm for $MULTI\_AND_m(f)$ using a completely different method, and for that reason we stay with inequality in total estimation. □

A similar result is obtained with a nondeterministic quantum query algorithm and the construction MULTI_OR.

**Theorem 4** *Let $Q_1$ be an exact quantum query algorithm that computes Boolean function f with k queries. Then a nondeterministic quantum query algorithm $Q_2$ exists that computes the function $MULTI\_OR_m(f)$ with the same k queries for all m.*

**Proof.** We use the same algorithm Q2 from the proof of Theorem 2.
Here we interpret the results of running Q2 as follows:

- If there exists at least one such $i \in \{1..m\}$ that $f(X_i) = 1$, then by Lemma 2 in a part of Q2 corresponding to $f(X_i)$ there will be $P_{f(X_i)}("1") = \frac{1}{m}$. The total probability of obtaining result "1" will be $p(f(X) = 1) > 0$ and in the view of nondeterministic algorithm definition we have $MULTI\_OR(X) = 1$.

- If for all $i \in \{1..m\}$ it is true that $f(X_i) = 1$, then in all Q2 parts $P_{f(X_i)}("0") = \frac{1}{m}$ and the total probability is $p(f(X) = 0) = 1$, so in the view of dual nondeterministic algorithm definition we have $MULTI\_OR(X) = 0$.

That completely agrees with the essence of function MULTI_OR; hence, Q2 computes this function as a nondeterministic algorithm. □

**Theorem 5** *For an arbitrary Boolean function f, $NQ_1(MULTI\_OR_m(f)) \leq Q_E(f)$.*

**Proof.** In a similar way as the proof of Theorem 3. □

In the next two theorems we generalize obtained results to make it possible to operate with compositions of arbitrary Boolean functions.

**Theorem 6** *Let $f_i$ be an arbitrary Boolean function. We consider a function $F = f_1 \wedge f_2 \wedge ... \wedge f_n$. Then a dual nondeterministic quantum query algorithm Q exists that computes F with $max(Q_E(f_1), Q_E(f_2), ..., Q_E(f_n))$ queries.*

**Proof.** We take all exact algorithms for $f_1, f_2, ..., f_n$ and run them in parallel, combining the queries.

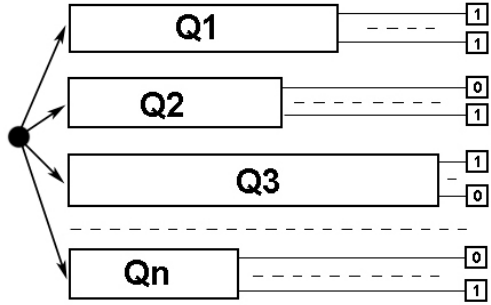

Fig. 4 Dual nondeterministic algorithm for $F = f_1 \wedge f_2 \wedge ... \wedge f_n$.

The proposition from Lemma 2 is true for algorithm $Q$; thus, it meet the properties of $F = f_1 \wedge f_2 \wedge ... \wedge f_n$. We are running exact algorithms and asking questions in parallel, so the complexity of the entire algorithm equals the largest number of queries of corresponding exact algorithms. □

**Theorem 7** *Let $f_i$ be an arbitrary Boolean function. We consider a function $F = f_1 \vee f_2 \vee ... \vee f_n$. Then a nondeterministic quantum query algorithm $Q$ exists that computes F with $\max(Q_E(f_1), Q_E(f_2), ..., Q_E(f_n))$ queries.*

**Proof.** Similar to the proof of Theorem 6.

## 4. Application

In this section, we describe several examples of how nondeterministic quantum query algorithms and their properties can be used to efficiently compute Boolean functions.

### 4.1. Demonstrative example

Now we will show how it is technically possible to construct a dual nondeterministic quantum algorithm for a composite function having exact quantum algorithms for sub-functions.

The following exact quantum query algorithms with two questions for functions $F_3(x_1, x_2, x_3) = \neg(x_1 \oplus x_2) \wedge (x_1 \oplus x_3)$ and $G_4(x_1, x_2, x_3, x_4) = (x_1 \oplus x_2) \wedge (x_3 \oplus x_4)$ were presented in [3].

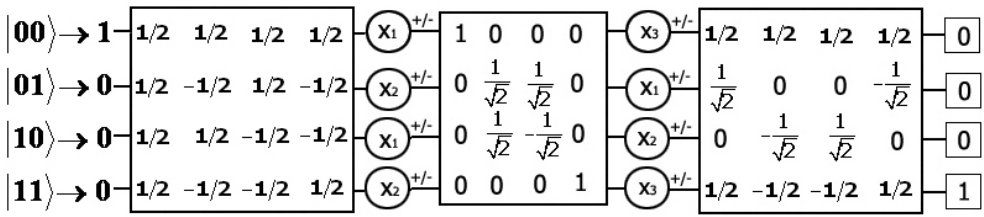

Fig. 5 Exact quantum query algorithm for function $F_3$.

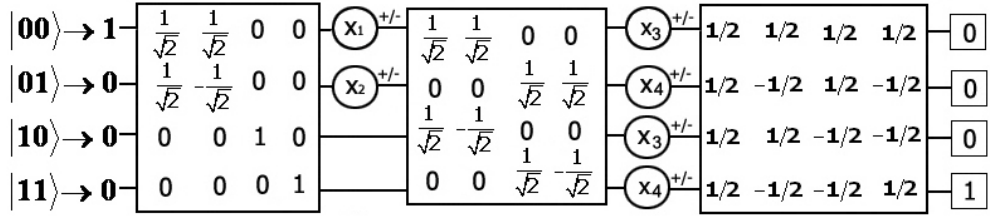

Fig. 6 Exact quantum query algorithm for function $G_4$.

Here, we consider a composite function that was obtained by combining $F_3$ and $G_4$:
$$H_7(x_1,x_2,x_3,x_4,x_5,x_6,x_7) = (\neg(x_1 \oplus x_2) \wedge (x_1 \oplus x_3)) \wedge ((x_4 \oplus x_5) \wedge (x_6 \oplus x_7))$$

The sensitivity of the base functions is $s(F_3) = 3$ and $s(G_4) = 4$. It is easy to show that the sensitivity of $H_7$ is equal to the number of variables, so we have $D(H_7) = 7$.

According to Theorem 6, a dual nondeterministic algorithm exists with only $\max(Q_E(F_3), Q_E(G_4)) = 2$ queries.

Using an approach similar to that one used in the proof of Theorem 2, we are able to completely describe a dual nondeterministic algorithm that computes $H_7$.

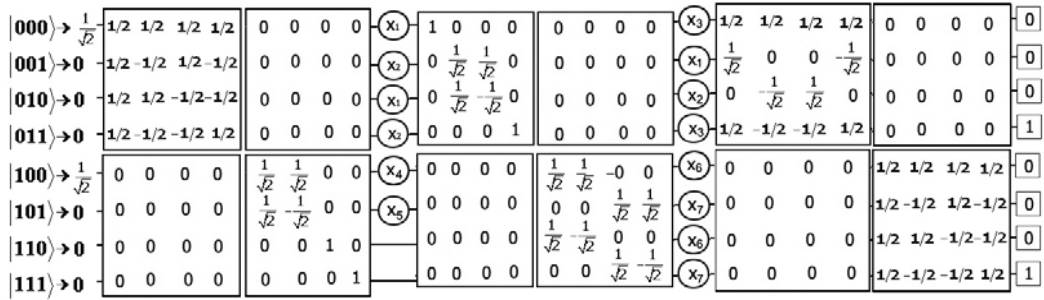

Fig. 7 Dual nondeterministic quantum query algorithm for $H_7$.

### 4.2. $Control_n$ function

We introduce a function and prove a gap between deterministic and dual nondeterministic quantum algorithm complexity.

Let $Control_n$ be a Boolean function of $n=2k-1$ variables:

$$Control_n(x_1, x_2, ..., x_k, x_{k+1}, ..., x_{2k-1}) = 1 \Leftrightarrow \begin{cases} x_{k+1} = x_1 \oplus x_2 \\ x_{k+2} = x_1 \oplus x_2 \oplus x_3 \\ .............................. \\ x_{2k-2} = x_1 \oplus x_2 \oplus ... \oplus x_{k-1} \\ x_{2k-1} = x_1 \oplus x_2 \oplus ... \oplus x_{k-1} \oplus x_k \end{cases}$$

The essence of the function is that in each accepting input $X$, such that $Control_n(X)=1$, values of the first $j$ bits control the value of $(k+j)$'s bit.

**Theorem 8** $D(Control_n) = n$

**Idea of Proof.** Use the sensitivity of a function on zero input $X=00..0$.

**Theorem 9** *There is a dual nondeterministic quantum algorithm Q that computes $Control_n$ with 2 queries for all n.*

**Proof.** First, we transform the equation system as follows:

$$\begin{cases} x_{k+1} = x_1 \oplus x_2 \\ x_{k+2} = x_1 \oplus x_2 \oplus x_3 \\ \dots \\ x_{2k-1} = x_1 \oplus x_2 \oplus \dots \oplus x_k \end{cases} \Rightarrow \begin{cases} x_{k+1} = x_1 \oplus x_2 \\ x_{k+2} = x_{k+1} \oplus x_3 \\ \dots \\ x_{2k-1} = x_{2k-2} \oplus x_k \end{cases} \Rightarrow \begin{cases} x_{k+1} \oplus x_1 \oplus x_2 = 0 \\ x_{k+2} \oplus x_{k+1} \oplus x_3 = 0 \\ \dots \\ x_{2k-1} \oplus x_{2k-2} \oplus x_k = 0 \end{cases}$$

We can rewrite the following statement:

$$\begin{cases} x_{k+1} \oplus x_1 \oplus x_2 = 0 \\ x_{k+2} \oplus x_{k+1} \oplus x_3 = 0 \\ \dots \\ x_{2k-1} \oplus x_{2k-2} \oplus x_k = 0 \end{cases} \Leftrightarrow Control_n(X) = 1$$

with equivalent logical formulas:

$$Control_n = \neg((x_{k+1} \oplus x_1 \oplus x_2) \vee (x_{k+2} \oplus x_{k+1} \oplus x_3) \vee \dots \vee (x_{2k-1} \oplus x_{2k-2} \oplus x_k))$$
$$Control_n = \neg(x_{k+1} \oplus x_1 \oplus x_2) \wedge \neg(x_{k+2} \oplus x_{k+1} \oplus x_3) \wedge \dots \wedge \neg(x_{2k-1} \oplus x_{2k-2} \oplus x_k)$$
$$Control_n = \neg(PARITY_3(X_1)) \wedge \neg(PARITY_3(X_2)) \wedge \dots \wedge \neg(PARITY_3(X_{k-1}))$$
$$Control_n = MULTI\_AND_{k-1}(\neg PARITY_3)$$

From Theorem 2 follows that a dual nondeterministic quantum algorithm $Q$ for a function $Control_n$ exists, such that $NQ_0(Q) = Q_E(\neg PARITY_3)$.

Figure 8 demonstrates a quantum exact algorithm with 2 questions for a function $\neg PARITY_3(X) = \neg(x_1 \oplus x_2 \oplus x_3)$.

Figure 9 demonstrates the structure of a complete dual nondeterministic algorithm computing $Control_n$.

Here $H$ is Hadamard gate $H = \frac{1}{\sqrt{2}}\begin{pmatrix} 1 & 1 \\ 1 & -1 \end{pmatrix}$

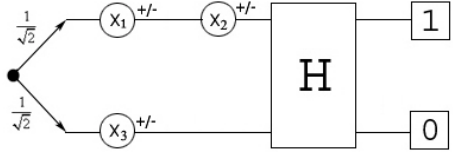

Fig. 8 Quantum exact algorithm for $\neg PARITY_3(X)$.

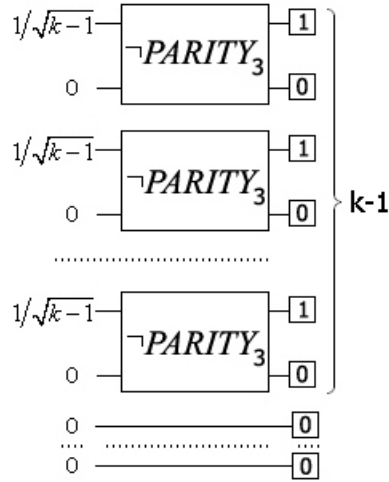

Fig. 9 Quantum dual nondeterministic algorithm for $Control_n$ with 2 queries.

## 5  Conclusion

In this paper, we studied nondeterministic quantum query algorithms, first introducing the new notion of a dual nondeterministic quantum query algorithm and then proving its relations to the complexity of nondeterministic and exact algorithms for several classes of Boolean functions. From the results, it becomes clear that the considered nondeterministic algorithms are a powerful and efficient model. We also introduce the new function $Control_n$ and describe a dual nondeterministic algorithm with 2 queries for all $n$.

The future direction of this research is to prove stronger relations to other types of query algorithms, for example, to exact algorithms of the same function, classical nondeterministic query algorithms and even classical deterministic. It would also be interesting to discover efficient quantum nondeterministic algorithms for specific functions, which would reveal large gaps between complexities of different kinds of algorithms.